\documentclass[conference]{IEEEtran} % [cite: 1]
\IEEEoverridecommandlockouts
\usepackage{cite}
\usepackage{amsmath,amssymb,amsfonts}
\usepackage{algorithmic}
\usepackage{graphicx}
\usepackage{textcomp}
\usepackage{xcolor}
\usepackage{booktabs}
\usepackage{multirow}
\usepackage{comment}
\usepackage{balance}
\usepackage{epsfig}
\usepackage[normalem]{ulem}
\usepackage{tabularx}
\usepackage{array}
\def\BibTeX{{\rm B\kern-.05em{\sc i\kern-.025em b}\kern-.08em
    T\kern-.1667em\lower.7ex\hbox{E}\kern-.125emX}}
\begin{document}

\title{An Asynchronous Delta Modulator for Spike Encoding in Event-Driven Brain-Machine Interface}

\author{
\IEEEauthorblockN{Kaushik Lakshmiramanan$^{\dagger}$, Vineeta Nair$^{\dagger}$, Ching-Yi Lin$^{\dagger}$, Sheng-Yu~Peng$^{*}$, and Sahil~Shah$^{\dagger}$}

\IEEEauthorblockA{$^{\dagger}$Department of Electrical and Computer Engineering, University of Maryland, College Park, MD, USA\\
$^{*}$Department of Electrical and Computer Engineering, National Yang Ming Chiao Tung University, Taiwan\\
Email: speng@nycu.edu.tw \quad sshah389@umd.edu }
}
\maketitle

\begin{abstract}
    This paper presents the design and implementation of an asynchronous delta modulator as a spike encoder for event-driven neural recording in a 65nm CMOS process. The proposed neuromorphic front-end converts analog signals into discrete, asynchronous ON and OFF spikes, effectively compressing continuous biopotentials into spike trains compatible with spiking neural networks (SNNs). Its asynchronous operation enables seamless integration with neuromorphic architectures for real-time decoding in closed-loop brain--machine interfaces (BMIs). Measurement results from silicon demonstrate an energy consumption of $60.73\,\mathrm{nJ/spike}$, an F1-score of $80\%$ compared to a behavioral model of the asynchronous delta modulator, and a compact pixel area of $73.45\,\mu\mathrm{m} \times 73.64\,\mu\mathrm{m}$.
\end{abstract}


\section{Introduction}
    The development of high-density intracortical microelectrode arrays (MEAs) over the past decades has significantly increased the volume of neural data acquired in neural interfaces \cite{liu2024flexible}. As electrode counts continue to scale, approximately doubling every 7 years, the acquisition and transmission of raw, high-bandwidth analog data from on-chip sensors to external processing units impose substantial demands on power and bandwidth \cite{stevenson_how_2011}. To address these challenges, a wide range of low-power analog front-end circuit topologies have been proposed \cite{bharucha_survey_2014, abdelgaliel_2_2024, liu_compact_2018}.
    
    A major contributor to power consumption in these analog front-ends is the analog-to-digital converter. However, neural decoders in brain-machine interfaces (BMIs) primarily rely on action potentials, or ``spikes,'' to decode neural activity into kinematic information \cite{gilja_high-performance_2012,christie_comparison_2015,wandelt_decoding_2022}. This observation motivates the use of alternative front-end architectures that directly encode analog signals into spikes, thereby eliminating the need for conventional data converters and reducing overall system power.
    
    Several approaches exist for encoding analog signals into spikes. Spike sorting techniques can extract neuron-specific activity from recorded signals, but they require computationally expensive algorithms and are often unsuitable for low-power, real-time systems \cite{zhang_spike_2023, han_l-sort_2025}. Alternatively, threshold-crossing methods based on noise-adaptive thresholds derived from the root mean square (RMS) voltage of the signal can be employed \cite{christie_comparison_2015}. While widely used in neural decoding systems \cite{1404138, 7101292}, these approaches require additional circuitry to estimate noise statistics. Simpler fixed-threshold methods reduce circuit complexity but are sensitive to variations in signal-to-noise ratio, leading to degraded performance under low-SNR conditions, as demonstrated in this work. To overcome these limitations, we propose an analog front-end based on delta modulation, a technique widely used in dynamic vision sensors for efficient event-based encoding \cite{4444573}.
    
\begin{figure}[h]
    \centering    \includegraphics[width=\linewidth]{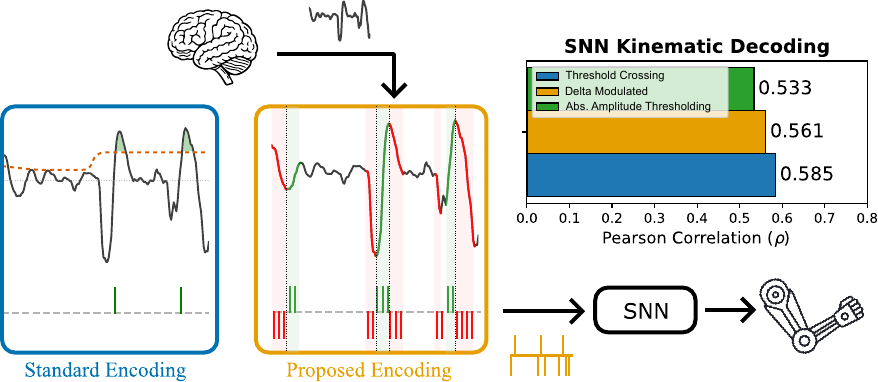}
    \caption{Proposed method in a closed-loop BMI and SNN decoding performance comparing spiked encoded using threshold crossing, Delta-Modulated (DM) and Absolute Amplitude Threshold spikes averaging X and Y velocities using Pearson correlation coefficient ($\rho$)}
    \label{fig:BMI_SNN}
    \vspace{-1mm}
\end{figure}

    The proposed front-end circuit asynchronously generates events when the input signal crosses predefined upward or downward thresholds relative to a running baseline. Figure~\ref{fig:BMI_SNN} illustrates the overall spike encoding mechanism. This encoding scheme captures changes in the input signal rather than its absolute amplitude, enabling efficient representation of neural activity. Furthermore, the proposed front-end supports asynchronous communication with address event representation (AER) circuits, facilitating seamless integration with event-driven neural decoders. Such decoders can further reduce system power while maintaining competitive decoding performance \cite{taeckens_spiking_2024,dethier_design_2013,mohan_architectural_2025}.

\section{Event-based Neural Decoding}

    While delta modulation offers advantages in area and power by eliminating explicit data conversion, its effectiveness for neural decoding remains an open question. In particular, it is important to evaluate whether delta-modulated spike representations can achieve performance comparable to conventional threshold-crossing methods.
    
    To this end, we evaluate the decoding performance of delta-modulated (DM) spikes using a spiking neural network (SNN) and compare it against threshold-crossing and absolute amplitude threshold-based encoding. The SNN incorporates an architecture similar to that of the model presented in \cite{taeckens_spiking_2024}. Threshold-crossing spikes were obtained from the Nonhuman Primate (Indy) reaching dataset \cite{ODoherty:2017}, which uses Blackrock Cerebus Neural Signal Processor's RMS-multiplier threshold crossing algorithm \cite{rizk_optimizing_2009}, serving as a reference for decoding performance. Target kinematics from this same dataset \cite{ODoherty:2017} were used as training labels, ensuring they were precisely temporally aligned with the corresponding neural spikes.
    
    As shown in Figure~\ref{fig:BMI_SNN}, the proposed asynchronous delta modulator-based encoding achieves competitive decoding performance. DM spikes achieve an average Pearson correlation coefficient ($\rho$) of 0.561, closely approaching the threshold-crossing performance of 0.585, while absolute amplitude threshold encoding shows slightly degraded performance. These results demonstrate delta modulation as an effective strategy in balancing the trade-offs between energy-efficient spike encoding and decoding accuracy.

\section{Asynchronous Delta Modulator}
\subsection{Background}
    Asynchronous delta modulators can be realized using a variety of circuit architectures. A well-established approach employs a capacitive feedback amplifier with a reset mechanism to integrate the input signal and generate asynchronous spikes using comparators when predefined voltage thresholds are exceeded \cite{sharifshazileh_electronic_2021}. Alternative time-domain and frequency-based encoding techniques utilize differential current-controlled ring oscillators, where the oscillation frequency is modulated by input-dependent bias currents \cite{9258945}. More recent designs have explored capacitive charge redistribution techniques inspired by successive approximation register (SAR) ADCs to achieve improved threshold control and linearity \cite{Elzakker-chargeRedist}. In addition, hybrid architectures combining SAR-based conversion with voltage-controlled oscillators within a $\Delta\Sigma$ loop have been proposed to leverage both coarse charge-domain and fine time-domain encoding \cite{sanyal_energy-efficient_2017}. In this work, a differencing circuit using a capacitive-feedback inverting amplifier combined with dual comparators for spike outputs is adopted. 

\subsection{Operating principle}
An asynchronous delta modulator operates by continuously monitoring the difference between an incoming signal $V_{\mathrm{in}}(t)$ and the last known value $V'{(t)}$, emitting a spike event only when this difference exceeds a fixed threshold ($\delta$) in either direction rather than sampling at uniform time intervals.
In the asynchronous delta modulator, spikes are generated as follows:
\begin{equation}
    V_\mathrm{on}:V_\mathrm{{in}}(t) - V'{(t)}\ge +\delta
\end{equation}
\begin{equation}
    V_\mathrm{off}:V_\mathrm{{in}}(t) - V'{(t)}\le -\delta
\end{equation}
\begin{figure}[htbp]
    \centering
    \includegraphics[width=\linewidth]{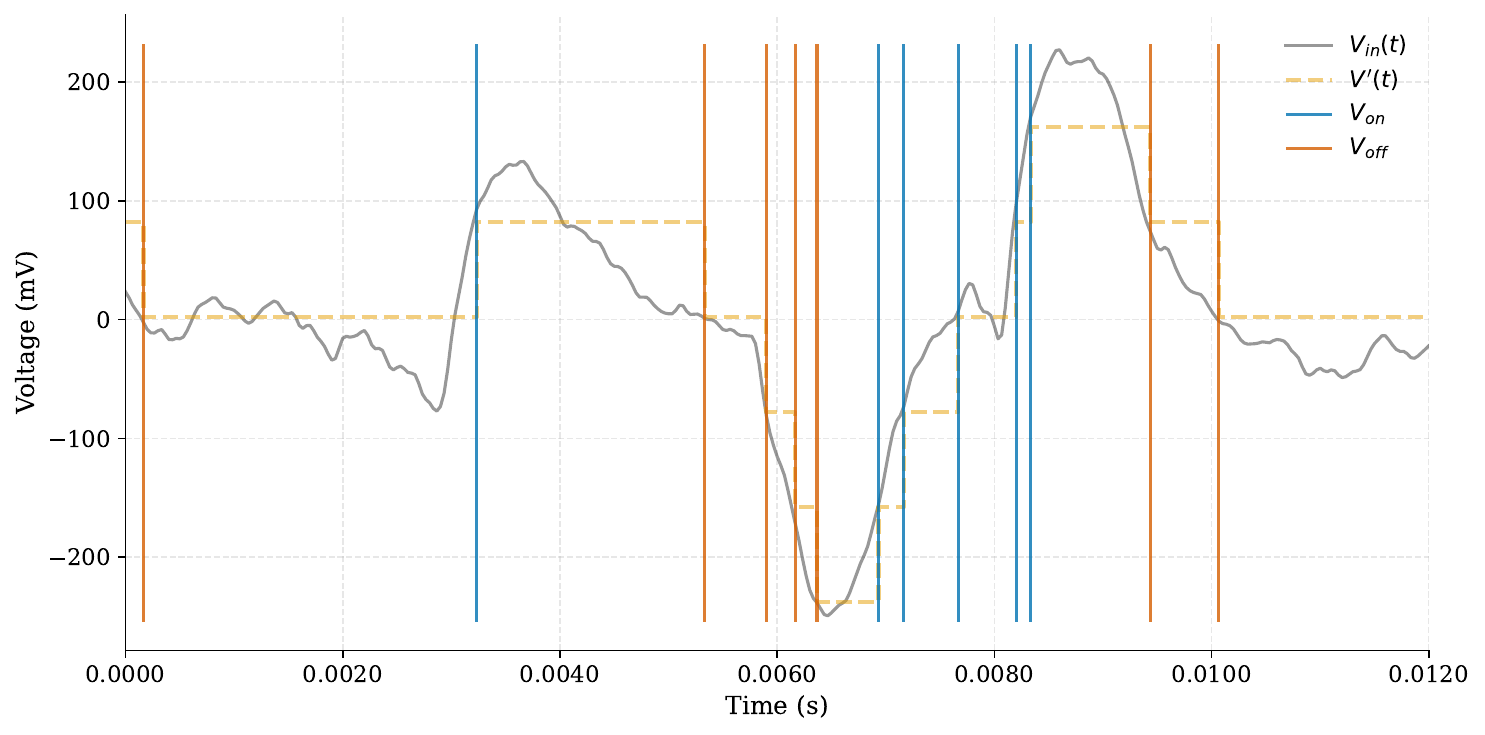}
    \caption{Single action potential encoded using the asynchronous delta modulator}
    \label{fig:single_action}
    \vspace{-1mm}
\end{figure}
 
where $V'{(t)}$ is the last known/triggered value. Figure~\ref{fig:single_action} depicts the nominal operation of the asynchronous delta modulator under a representative input signal, where positive and negative spike events are asynchronously generated corresponding to upward and downward threshold crossings, respectively. This asynchronous, level crossing behavior ensures the output spike rate scales naturally with the rate of change of the input signal, and rapidly varying signal amplitudes produce dense spike trains while quiescent periods yield silence, making the asynchronous delta modulator inherently well matched to the sparse, temporally coded representations used in neuromorphic processing pipelines.

\subsection{Circuit Topology}

\begin{figure}[t]
    \centering
    \includegraphics[width=\linewidth]{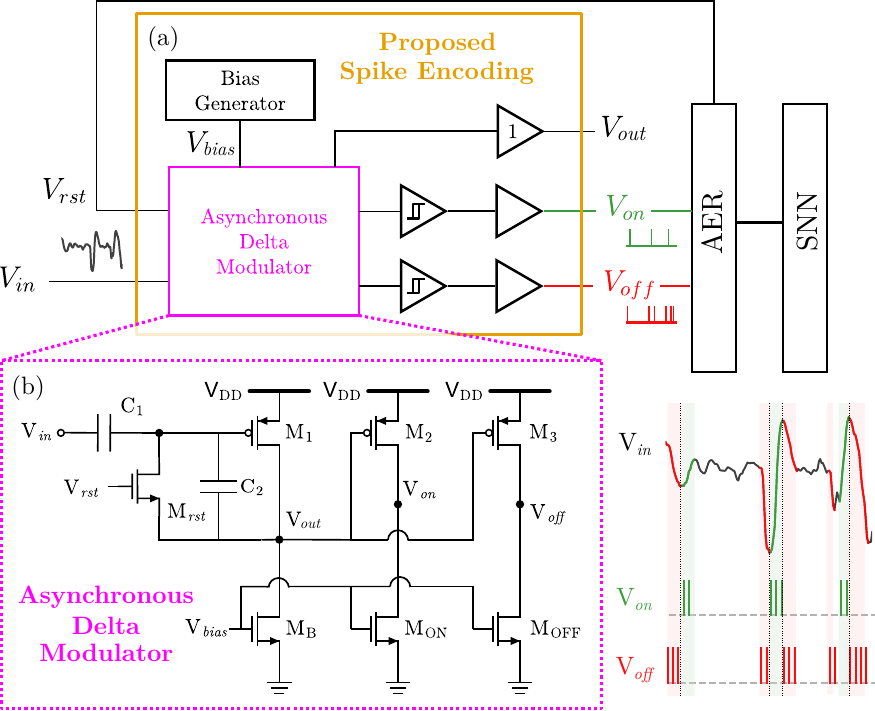}
    \caption{(a) System-level block diagram of the asynchronous delta modulator interfacing with the AER communication block (b) Transistor-level schematic of the Asynchronous Delta Modulator showing the capacitive coupling, input stage, and comparator stages}
    \label{fig:circuit_dm}
    \vspace{-3mm}
\end{figure}

The system architecture of the proposed event-driven front-end is illustrated in Figure~\ref{fig:circuit_dm}(a). At its core, the asynchronous delta modulator receives an amplified analog signal from a Low-Noise Amplifier (LNA) $V_{\mathrm{in}}$ and evaluates it against the threshold set by the bias generator $V_{\mathrm{bias}}$. Any deviation beyond the defined threshold triggers a comparator stage, the output of which triggers two Schmitt triggers to produce full-swing digital pulses. This encoding yields two distinct spike sequences for positive $V_\mathrm{{on}}$ and negative transitions $V_\mathrm{{off}}$, respectively. The resulting asynchronous pulse streams can be routed to an Address Event Representation (AER) block for multiplexed, low-latency spike transmission to downstream processing or decoding stages. 
% An additional buffer stage delivers an isolated output of $V_{diff}$ at a dedicated node for circuit characterization and functional verification during testing.

The transistor-level circuit of the asynchronous delta modulator is depicted in Figure~\ref{fig:circuit_dm}(b). The topology is a capacitively-coupled, single-ended inverting amplifier paired with a reset mechanism for asynchronous sampling. The analog neural input \textit{$V_{\mathrm{in}}$} is AC-coupled through capacitor $C_\mathrm{1}$, which suppresses DC offsets inherited from any preceding amplifier stages. Capacitor $C_2$ serves as the feedback element with a reset transistor $M_{\mathrm{rst}}$ connected in parallel to it. $M_{\mathrm{rst}}$, gated by $V_\mathrm{{rst}}$ and triggered by the AER or external logic, shorts the input with the output of the common-source amplifier following each spike event. The delay of the reset pulse determines the output spike width, while the reset pulse width determines the refractory period of the output spikes. This allows control over the frequency of output spikes and reduces interference from erratic signals. A Beta-Multiplier Reference provides the required bias voltage for transistors $M_\mathrm{B}$, $M_{\mathrm{ON}}$, and $M_{\mathrm{OFF}}$. Transistor $M_\mathrm{B}$, biased by $V_{\mathrm{bias}}$, establishes the operating current of the inverting amplifier, which also defines the steady-state reference value to which the output returns during reset. Transistors $M_1$, $M_2$, and $M_3$ constitute the amplification and tracking of the voltage variations at the $V_{\mathrm{out}}$ node, where, for a small time interval $dt$, the voltage change is given by:
\begin{equation}
    \label{diffVout}
    \frac{dV_{\mathrm{out}}}{dt} = -A \frac{dV_{\mathrm{in}}}{dt}
\end{equation}
where A = ${C_1/C_2}$ is the differencing amplifier gain. 
After a time interval $\Delta$t, the output voltage is given by integrating Eq. \ref{diffVout}.
\begin{equation}
    \label{Vout}
    V_{\mathrm{out}} = -A \int_{t}^{t+\Delta t} dV_{\mathrm{in}} = -A\cdot\Delta V_{\mathrm{in}}
\end{equation}
If the output voltage of the amplifier exceeds the thresholds set by the bias, the comparators $M_\mathrm{{ON}}$ or $M_\mathrm{{OFF}}$ pull their respective nodes low or high, respectively, until the reset pulse is triggered. The spikes are generated from the comparator outputs by triggering the subsequent Schmitt triggers illustrated in Figure~\ref{fig:circuit_dm}(b) to produce the final spike outputs $V_{\mathrm{on}}$ and $V_{\mathrm{off}}$. The spike outputs are used to generate a reset signal $V_{\mathrm{rst}}$, which brings the output node of the amplifier back to the reference value to start the next integration cycle based on the input. 

Tuning the refractory period to a higher value helps protect against high-frequency noise while giving up on the accuracy. In a closed-loop system, the spikes would be transmitted asynchronously from a multi-electrode array using Address-event representation (AER) and then to the SNN for further decoding.

\begin{figure}[b]
    \centering
    \includegraphics[width=\linewidth]{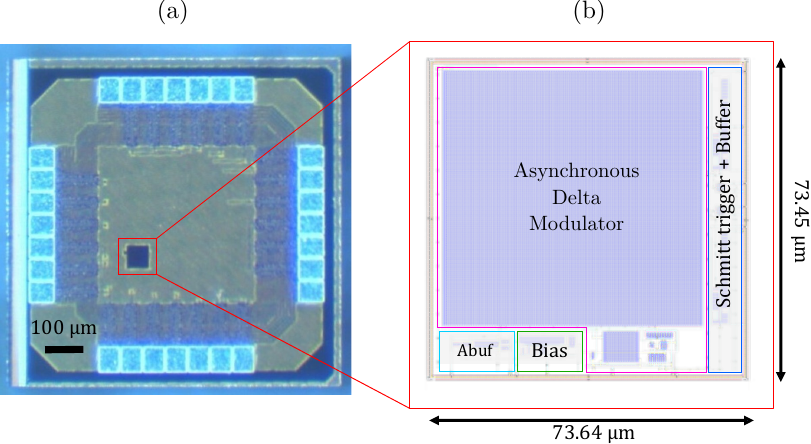}
    \caption{(a) Taped-out chip (b) Enlarged view of the asynchronous delta modulator, highlighting the core, bias, buffers, and Schmitt triggers}
    \label{fig:chip_layout}
    \vspace{-3mm}
\end{figure}

Figure~\ref{fig:chip_layout}(a) presents the complete chip floorplan alongside a detailed inset of the asynchronous delta modulator's layout. As highlighted in Figure~\ref{fig:chip_layout}(b), the core circuitry, especially the coupling and feedback capacitors ($C_{\mathrm{1}}$ and $C_{\mathrm{2}}$), occupies the majority of the area.

\begin{table}[htbp]
    \caption{Asynchronous Delta Modulator Hardware Measurement}
    \label{tab:adm_specs}
    \centering
    \renewcommand{\arraystretch}{1.5} % Row height
    \setlength{\tabcolsep}{15pt} % Column separation
    % \begin{tabular}{l l}
    \begin{tabular}{l l}
        \hline
        \textbf{Parameter} & \textbf{Value} \\
        \hline
            CMOS Process &  65nm \\
            Supply Voltage ($V_{\mathrm{DD}}$) & 1.2 V \\
            Pixel Area & 0.0054 $mm^2$ \\
            % Signal Bandwidth &  80 Hz–8 kHz \\
            Mid-band Gain & 12.14 dB \\
            Low Cut-off (-3dB) &  80 Hz \\
            High Cut-off (-3dB) &  8 kHz \\
            Input-referred Noise & 14.990 $\mu V_\mathrm{rms}$ \\
            Power (Dynamic) & 12.145 $\mu W$ \\
            Energy / Spike &  60.7281 nJ/spike \\
        \hline
    \end{tabular}
\end{table}

\section{Experimental Results}
 \begin{figure}[htbp]
        \centering
        \includegraphics[width=\linewidth]{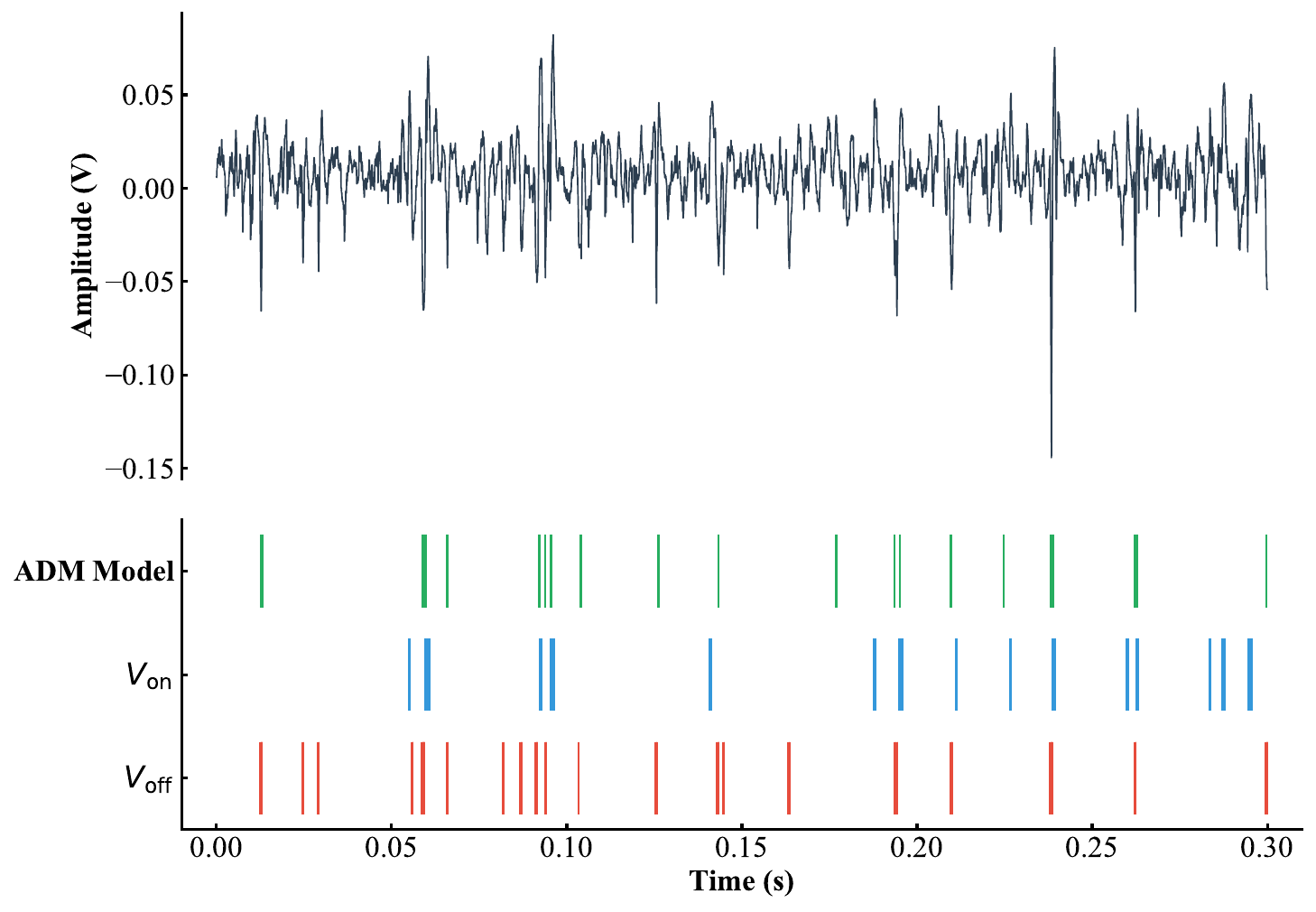}
        \caption{Transient response showing the analog input waveform ($V_{\mathrm{in}}$) for a 30$ms$ duration and the resulting asynchronous $V_{\mathrm{on}}$ (blue) and $V_{\mathrm{off}}$ (red) spike raster compared against the spikes generated using a behavioral model (green)}
        \label{fig:spike_output2}
        \vspace{-4mm}
    \end{figure}
    
% \subsection{Transient Characterization and SNR analysis}
The hardware measurement results of the fabricated asynchronous delta modulator chip are summarized in Table \ref{tab:adm_specs}. The mid-band gain and the cut-off frequencies (-3dB) are calculated by sweeping an AC input signal across a frequency range of 0.1Hz to 1 GHz and measuring the output magnitude. The frequency bandwidth allows the capture of action potentials, which have a band of about 250 Hz-5 kHz \cite{Harrison_2009}. An additional analog buffer (Abuf in Figure~\ref{fig:chip_layout}(b)) provides an output of $V_{\mathrm{out}}$ at a dedicated node for circuit characterization and functional verification during testing.

To analyze the input-referred noise, the asynchronous delta modulator's input is grounded, and the output is captured with a spectrum analyzer. An SR560 preamplifier is used to amplify the signal above the instrument's noise floor, and the measured output noise is then divided by the combined gain of the preamplifier and the chip's common-source amplifier. This yields a noise spectral density of $91.88 nV/\sqrt{Hz}$. Integrating the noise spectral density across the operational bandwidth yields a total input-referred noise of 14.990 $\mu V_{\mathrm{rms}}$. 

The transient behavior of the circuit is demonstrated in Figure~\ref{fig:spike_output2}, where the spike outputs for a time-varying input signal ($V_{\mathrm{in}}$) are measured from the chip for a duration of 2 seconds. The input is an intracortical signal sampled at 30KSPS. In this work, the reset is generated externally, with a delay of 0.1$ms$ and a refractory period of 1$ms$. Since the positive spikes are inverted at the Schmitt trigger's output, they are flipped digitally for use with the external reset logic and representation. Positive excursions in the input signal successfully trigger the spikes which correspond to the $V_{\mathrm{on}}$ events, and negative excursions trigger the spikes corresponding to the $V_{\mathrm{off}}$ events. These spike events are compared against the behavioral model of the asynchronous delta modulator to calculate the accuracy and tune the thresholds accordingly.

\begin{figure}[h]
    \centering   \includegraphics[width=\linewidth]{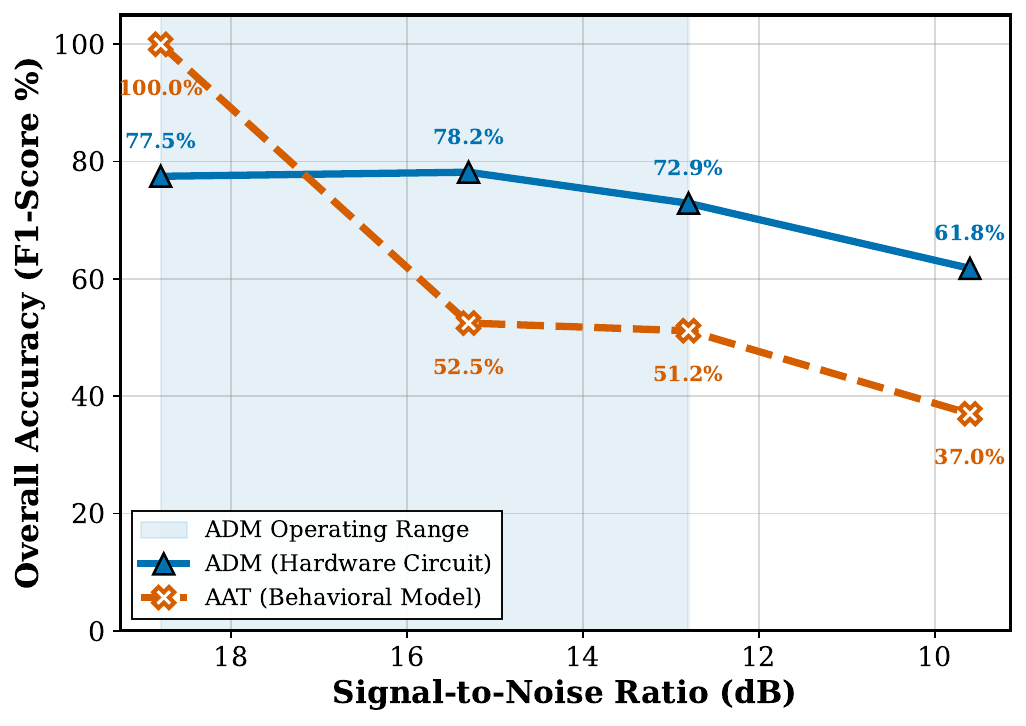}
    \caption{Robustness of the asynchronous delta modulator with increasing noise compared against the performance against the behavioral model of absolute amplitude modulation}
    \label{fig:SNR_Accuracy}
    \vspace{-3mm}
\end{figure}

To evaluate the robustness of the asynchronous delta modulator, an analysis was conducted by injecting the intracortical input signals with Additive White Gaussian noise (AWGN) at varying SNR levels, simulating the degraded recording conditions commonly encountered in chronic implant scenarios. Figure~\ref{fig:SNR_Accuracy} illustrates the performance of the hardware against a behavioral absolute amplitude thresholding model across a range of SNR. The baseline noise floor of the recorded signals was estimated at approximately 19 $dB$, quantified using the Median Amplitude Deviation (MAD) of the input signal. Zero-mean Gaussian noise was injected at four progressive levels: 1$\times$, 1.5$\times$, 2$\times$, and 4$\times$ the Median absolute value of the signal (22.84 $mV$), spanning a range of approximately 10.4\% to 41.5\% of the 220 $mV$ peak signal amplitude.

 %By adding white Gaussian noise to the baseline signal, the input signal's SNR was degraded, and the corresponding spike outputs were measured to calculate the accuracy.  
With increasing noise level, the behavioral absolute amplitude thresholding model exhibits a sharp deterioration in performance, with its F1-score collapsing from its perfect accuracy to approximately 50\% around 15 $dB$. On the contrary, the hardware asynchronous delta modulator demonstrates robustness, maintaining a stable F1-score of an average of 76.2\% over a band of 19 to 15 $dB$. A consistent trend observed in both encoders is an increase in sensitivity (recall) at lower SNRs, as elevated noise levels trigger additional threshold crossings that manifest as false positives, partially stabilizing the F1-score at higher noise floors. However, the asynchronous delta modulator sustains its overall performance through higher precision, reflecting its ability to accurately capture true spike events while suppressing spurious noise. This robustness under degraded SNR conditions is a direct consequence of the inherent noise-shaping properties of asynchronous delta modulation, which shifts the quantization noise from the signal band to higher, out-of-band frequencies \cite{1010643}. 

\section{Conclusion}
% This work presents the implementation and experimental validation of an ADM for spike encoding a continuous-time analog signal to digital spike trains. The ADM converts continuous intra-cortical analog signals into sparse, neural activity-dependent spike representations natively compatible with spiking neural network architectures, eliminating the representational mismatch between analog front-end acquisition and neuromorphic inference. 

% Robustness evaluation against additive noise spanning approximately 10.4\% to 41.5\% of the peak signal amplitude demonstrated the encoder's resilience under degraded recording conditions, a critical requirement for chronic implant scenarios. Furthermore, behavioral decoding experiments conducted on the NHP reach-to-grasp dataset confirmed that the ADM-encoded spike trains preserve sufficient task-relevant neural information, achieving a Pearson correlation on par with that of computationally intensive ground-truth spike-sorted signals. Collectively, these results validate the proposed ADM as a hardware-efficient, noise-resilient neuromorphic front-end capable of bridging biological neural signals and spike-based computational frameworks in real-time closed-loop BMI systems.

This work presents the implementation and experimental validation of an Asynchronous Delta Modulator designed to encode continuous-time analog signals into digital spike trains. By converting intra-cortical signals into sparse, activity-dependent representations, the asynchronous delta modulator eliminates the representational mismatch between analog acquisition and Spiking Neural Network (SNN) architectures.
Robustness evaluations against additive noise (ranging from 10.4\% to 41.5\% of peak amplitude) demonstrate the encoder’s stability under the degraded recording conditions typical of chronic implants. These results validate the proposed asynchronous delta modulator as a neuromorphic front-end capable of bridging biological signals and spike-based computational frameworks for real-time, closed-loop BMI systems.

\balance

\bibliographystyle{IEEEtran}
% \bibliography{references.bib,references_Shah}
\bibliography{Main}

\end{document}